\documentclass[a4paper,11pt]{article}
\usepackage{pos}

\title{$b \to s\gamma$, electron EDM and electroweak baryogenesis: 
  \\ a study in general 2HDM}
 \ShortTitle{$b \to s\gamma$, eEDM and EWBG: a study in g2HDM}

\author{George Wei-Shu Hou}

\affiliation{National Taiwan University, 
  1 Roosevelt Road, Sec. 4, Taipei 10617, Taiwan, R.O.C.}

\emailAdd{wshou@phys.ntu.edu.tw}

\abstract{
We study inclusive $b \to s\gamma$ decay in the context of electron EDM 
and baryogenesis. The {\it general} 2HDM (i.e.~without $Z_2$) that 
possesses an extra set of Yukawa matrices can drive electroweak baryogensis 
via $\lambda_t {\rm Im}\rho_{tt}$, where $\rho_{tt}$ is the extra diagonal 
top Yukawa coupling, with the $e$EDM constraint evaded by an exquisite 
flavor cancellation mechanism. We touch upon the current status of direct 
search for exotic $H$, $A$ and $H^+$ scalars at the LHC, while for the 
plethora of flavor observables in $g$2HDM, we focus on $b\to s\gamma$,
pointing out {\it chiral} enhancement of a $\rho_{tt}\rho_{bb}$ effect in 
{\it g}2HDM, which can bring in a CPV phase. We first explore the inclusive 
${\cal B}(b \to s\gamma)$ rate, then showcase the progress that Belle~II can 
make in the future, illustrating a potential 3$\sigma$ effect for the inclusive 
$B^+ \to X_s^+\gamma$ vs $B^0 \to X_s^0\gamma$ CPV rate difference, 
$\Delta A_{\rm CP}$. Especially if $e$EDM emerges swiftly, perhaps one 
should pursue further upgrade beyond Belle~II.
}

\FullConference{
EPS-HEP, 
 21-25 August 2023, 
Hamburg, Germany\\}

\begin{document}
\maketitle

\section{Introduction: tension, \& {\boldmath $t \to ch$}}

A three-way tension exists in present day particle physics: to have the more testable 
electroweak baryogenesis (EWBG), one needs very large $CP$ violation 
(CPV) beyond the Standard Model (BSM), which runs into tension with No New 
Physics (${\cal NNP}$) observed so far at the LHC.

Second, be it ACME~\cite{ACME:2018yjb} or JILA~\cite{Roussy:2022cmp}, 
electron EDM at the L.E. precision frontier provide sanity checks on large 
BSM-CPV, with stringent bound of $|d_e| < 0.41 \times 10^{-29} e\,{\rm
 cm}$~\cite{Roussy:2022cmp}. Finally, it can be said that these ``tabletop'' 
experiments are competing head-on with the behemoth LHC.

But it is fair to say that EWBG ought to be pursued while LHC is still running! 

We advocate~\cite{Hou:2021wjj} the {\it general} two Higgs doublet model 
({\it g}2HDM); unlike the usual 2HDM with $Z_2$ symmetry, it has a second 
set of Yukawa matrices that possess flavor changing neutral couplings (FCNC). 
With no theorem against a second Higgs, 2HDM should be a no-brainer, but we 
move the well-known NFC condition of Glashow-Weinberg off its pedestal, as 
it is nothing but {\it ad hoc}.

A hallmark of {\it g}2HDM would be $t \to ch$~\cite{Hou:1991un}, with $h$ the 
observed SM-like Higgs boson. Remarkably, besides flavor-hierarchies, {\it Nature} 
seems to throw in the emergent {\it alignment} phenomenon (small $h$-$H$ mixing, 
i.e. $c_\gamma \equiv \cos\gamma$ is small, with $H$ the exotic $CP$-even scalar) to 
protect this decay, with current limit at $0.00043$~\cite{ATLAS:2023ujo}.
The combined coupling $\rho_{tc}c_\gamma$ now barely allows $\rho_{tc}$ at 
${\cal O}(1)$.

\section{General 2HDM: EWBG \& $e$EDM} 

We do not show the {\it g}2HDM Higgs potential, but note that the convention is to 
take the Higgs basis where only one doublet, $\Phi$, gives v.e.v., as without $Z_2$, 
one cannot distinguish $\Phi$ from $\Phi'$. A minimization condition cancels the 
soft $\Phi^\dag\Phi'$ term against half the $|\Phi|^2\Phi^\dag\Phi'$ term, with the 
latter $\eta_6/2$ quartic coupling playing the unique role of $\Phi$-$\Phi'$ mixing.

One advantage of {\it g}2HDM is that {\cal O}(1) quartics $\eta_i$ can lead 
to~\cite{Kanemura:2004ch} 1$^{\rm st}$ order phase transition (1$^{\rm st}\cal O$PhT), 
one of the Sakharov conditions. It was then argued~\cite{Hou:2017hiw} that the exotic 
$H/A/H^+$ bosons would likely be sub-TeV in mass, ripe for search at the LHC. The 
Yukawa couplings are 
\begin{align}
    - \frac{1}{\sqrt{2}} \sum_{f = u, d, \ell} 
 \bar f_{i} &\Big[\big(-\lambda^f_i \delta_{ij} s_\gamma + \rho^f_{ij} c_\gamma\big) h
  + \big(\lambda^f_i \delta_{ij} c_\gamma + \rho^f_{ij} s_\gamma\big)H
  - i\,{\rm sgn}(Q_f) \rho^f_{ij} A\Big]  R\, f_{j} \notag\\
  - \bar{u}_i&\left[(V\rho^d)_{ij} R-(\rho^{u\dagger}V)_{ij} L\right]d_j H^+ 
 - \bar{\nu}_i\rho^\ell_{ij} R \, \ell_j H^+
 +{h.c.},
\end{align}
with generation indices $i$, $j$ summed over, $L, R = (1\mp\gamma_5)/2$ are 
projections, $V$ the CKM matrix, with the lepton matrix taken as unity. With 
$c_\gamma$ small (with $s_\gamma \to -1$), the $h$ couplings are close to 
diagonal, while extra $\rho^f$ matrices are associated more with exotic scalars.
An interesting aspect of $H^+$ couplings in Eq.~(1) is that, by expanding 
$\rho^{u\dag}V$, one finds $H^+ \to c\bar b$, $t\bar b$ couplings are 
$\rho_{tc}V_{tb}$ and $\rho_{tt}V_{tb}$, resp., so unlike in 2HDM-II, 
$H^+ \to c\bar b$ is not CKM suppressed~\cite{Ghosh:2019exx}.

\begin{figure*}[t!]
\center
\includegraphics[width=5.47cm,height=3.6cm]{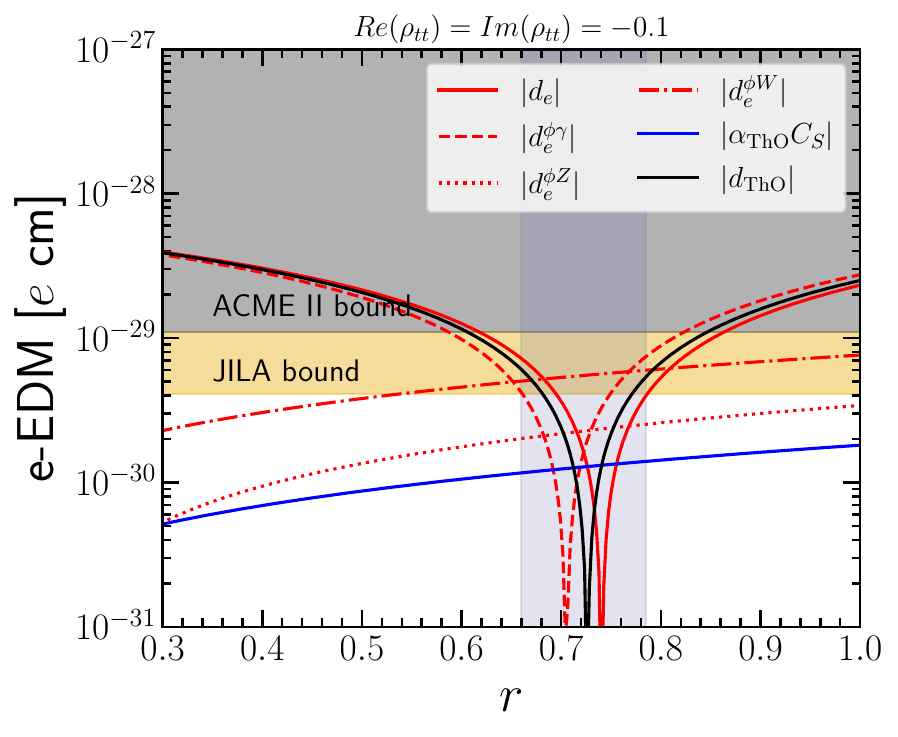}\includegraphics[width=4.78cm,height=3.6cm]{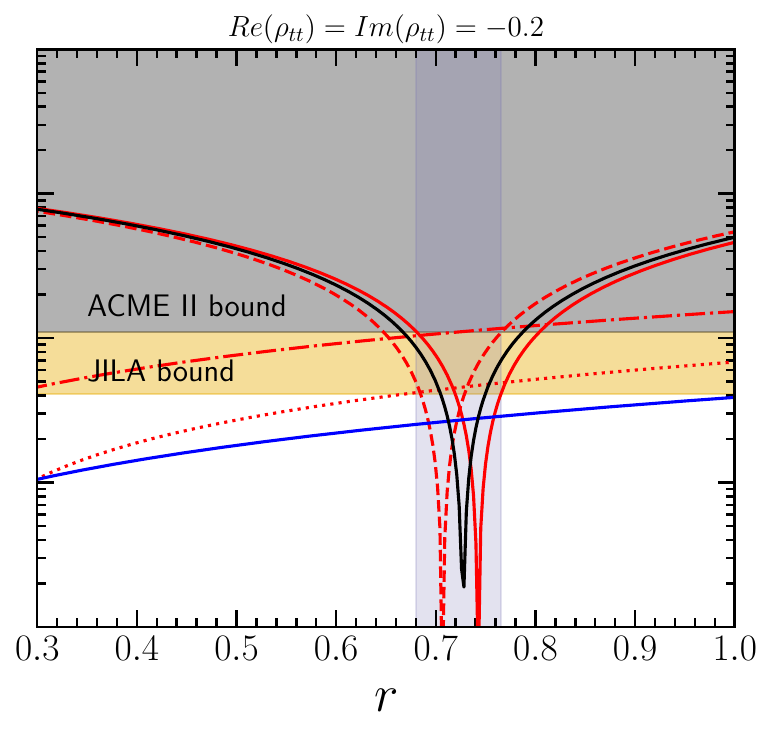}
\includegraphics[width=4.78cm,height=3.6cm]{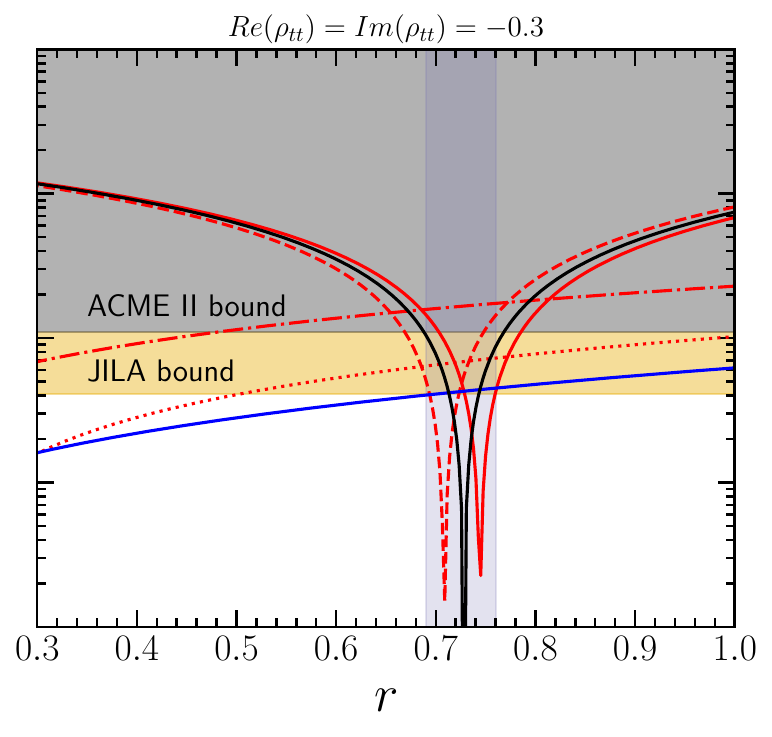}
\vskip-0.05cm
\caption{
 Cancellation of eEDM~\cite{Fuyuto:2019svr} for $m_H, m_A, m_{H^+} =$
 500 GeV, $c_\gamma = 0.1$, and $|\rho_{tt}| \cong 0.14$, 0.28, 0.42.
}
\label{fig:eEDM}
\end{figure*}

One highlight of {\it g}2HDM is that $\lambda_t {\rm Im} \rho_{tt}$ can drive 
EWBG robustly~\cite{Fuyuto:2017ewj}, while the leading two-loop Barr-Zee diagram 
for {\it e}EDM can be exquisitely cancelled~\cite{Fuyuto:2019svr} by flavor hierarchies 
quite effectively: the $\rho$ matrices {\it know} the SM flavor structure. 
In a recent paper~\cite{Hou:2023kho,TeunissenTalk} we revisited this cancellation 
mechanism and explored the larger range of 
\begin{align}
 {\rm Re}\rho_{tt} = {\rm Im}\rho_{tt} = -0.1, -0.2, -0.3, 
\end{align}
as illustrated in Fig.~1. Although the allowed range in loop function $r$ shrinks 
for larger $|\rho_{tt}|$, it was stressed that especially for $|\rho_{tt}|  \simeq 0.42$, 
{\it e}EDM could emerge rather swiftly. It could then be followed by n2EDM at 
PSI~\cite{n2EDM:2021yah} in {\it just} a few years. But a ``new" cancellation mechanism~\cite{Hou:2023kho} was also illustrated for $n$EDM itself through 
the unknown phase of $\rho_{uu}$, and the {\it second-whammy} from $n$EDM 
could take up to even two decades to pan out, but is still quite exciting.

\section{\boldmath $H$, $A$, $H^+$ Search \& Flavor Frontiers}

Sub-TeV exotic scalars should clearly be searched for at the LHC, while the 
{\it flavor} frontier is also quite promising. Here we limit ourselves to 
$b \to s\gamma$ for the latter, but let us first address a question: If there 
is a second Higgs doublet with a host of quartic and Yukawa couplings, 

\vskip0.13cm
\centerline{``Why is {\it g}2HDM hiding so well?''}
\vskip0.05cm

\noindent Exploring this question, we {\it guessed}~\cite{Hou:2020itz} a ``rule of thumb'' 
for flavor control:
\begin{align}
 \rho_{ii} \lesssim {\cal O}(\lambda_i), \ \rho_{1i} \lesssim {\cal O}(\lambda_1), \ 
 \rho_{3j} \lesssim {\cal O}(\lambda_3) \ (j \neq 1),
\label{eq:thumb}
\end{align}
which echoes the {\it e}EDM cancellation mechanism: the $\rho^f$ ($f = u, d, \ell$) 
matrices ``{\it know}'' the SM flavor structure, which roughly addresses the question
we raised above.

\vskip0.15cm
\noindent{\it Leading Search Modes at the LHC.---}
With $t \to ch$ suppressed by $c_\gamma$, it is natural to pursue $H/A/H^+$ 
direct production, gaining a $s_\gamma \to -1$ factor. 
The leading processes are~\cite{Kohda:2017fkn} 
\begin{align}
cg \to tH/tA \to tt\bar c, tt\bar t, 
\end{align}
where the second step follows from $H/A \to t\bar c$, $t\bar t$ decay via 
$\rho_{tc}$ and $\rho_{tt}$ couplings. Noting that $H^+ \to c\bar b$, 
$t\bar b$ have the same $V_{tb}$ factor and hence on equal footing, it 
was found that~\cite{Ghosh:2019exx} 
\begin{align}
cg \to bH^+ \to bt\bar b,
\end{align}
may be more efficient, with production via $\rho_{tc}$ and $H^+$ decay 
via $\rho_{tt}$, bypassing the heavy associated top in Eq.~(4).
Both ATLAS~\cite{ATLAS:2023tlp} and CMS~\cite{CMS:2023fod} have 
studied process (4), where the CMS study limits to $tt\bar c$ final state only. No 
signal is found so far, which might be expected, and one awaits adding Run~3 
data. Process (5) apparently has not been studied yet, which we look forward to.

\section{Chiral-enhanced $b \to s\gamma$ Observables}

The $b \to s\gamma$ process as a probe of $H^+$ effects in 2HDM-I and 
II~\cite{Hou:1987kf} still provides the best respective bounds on $m_{H^+}$. 
Here we are interested in the effects in {\it g}2HDM~\cite{Hou:2023ran}, 
finding
\begin{align}
       \delta C_{7, 8}^{(0)}(\mu)
  =  \frac{|\rho_{tt}|^2}{3|\lambda_t|^2} F_{7, 8}^{(1)}(x_t)
    - \frac{\rho_{tt}\rho_{bb}}{\lambda_t\lambda_b}F_{7, 8}^{(2)}(x_t),
\label{eq:C78}
\end{align}
where $x_t=m_t(\mu)^2/m_{H^+}^2$ is at heavy scale $\mu$. The $\lambda_{t,b}$ 
couplings in the denominators are actually {\it masses}, hence the second term is
$m_t/m_b$-enhanced, as it is rooted in {\it chiral} $H^+$  couplings. Since 
$|\rho_{bb}| \sim 0.1$ can also drive EWBG, it can be probed by $b \to s\gamma$ 
via chiral enhancement as well.

Defining $\phi \equiv {\rm arg}\,\rho_{tt}\rho_{bb} = \phi_{tt} + \phi_{bb}$, we use 
\texttt{Flavio\;v2.4.0}~\cite{Straub:2018kue} to estimate $b\to s \gamma$ observables, 
with WCs at heavy scale $\mu \sim m_{H^+}$ evolved down to the physical scale. 
We consider the well-measured inclusive ${\cal B}(B \to X_s\gamma)$, and the 
inclusive CPV difference~\cite{Benzke:2010tq},
\begin{align}
\Delta A_{\rm CP}(b\to s\gamma) \equiv A_{\rm CP}(B^+ \to X_s^+ \gamma)
 - A_{\rm CP}(B^0 \to X_s^0 \gamma),
\end{align}
for the future, where the projection of Belle~II physics book is used. We use 
HFLAV'21 for current data. Note that the error in ${\cal B}(B \to X_s\gamma)$ 
improves by less than a factor of two with full Belle II data. What's notable then 
is the projected order of magnitude improvement in $\Delta A_{\rm CP}$. 
Although the current sign from HFLAV'21 is insignificant, 
the numerics of Fig.~2 corresponds to the HFLAV'21 value in the Table below.

\begin{table}[h]
\centering
  \begin{tabular}{l|ccc}
   Observable & HFLAV'21  & Belle II ($5 \text{ ab}^{-1}$) & Belle II ($50 \text{ ab}^{-1}$)\\
    \hline
    $\mathcal{B}(B \to X_s \gamma)$& $(3.49 \pm 0.19)\times 10^{-4}$ & $(3.49 \pm 0.14)\times 10^{-4}$ & $(3.49 \pm 0.11)\times 10^{-4}$\\
    $\Delta A_{\rm CP}$ & $(3.70 \pm 2.80) \times 10^{-2} $ & $(0 \pm 0.98) \times 10^{-2} $ & $(0 \pm 0.30) \times 10^{-2} $
  \end{tabular}
\end{table}
\vskip-0.2cm
\begin{figure}[h]
     \includegraphics[width=0.29\textheight]{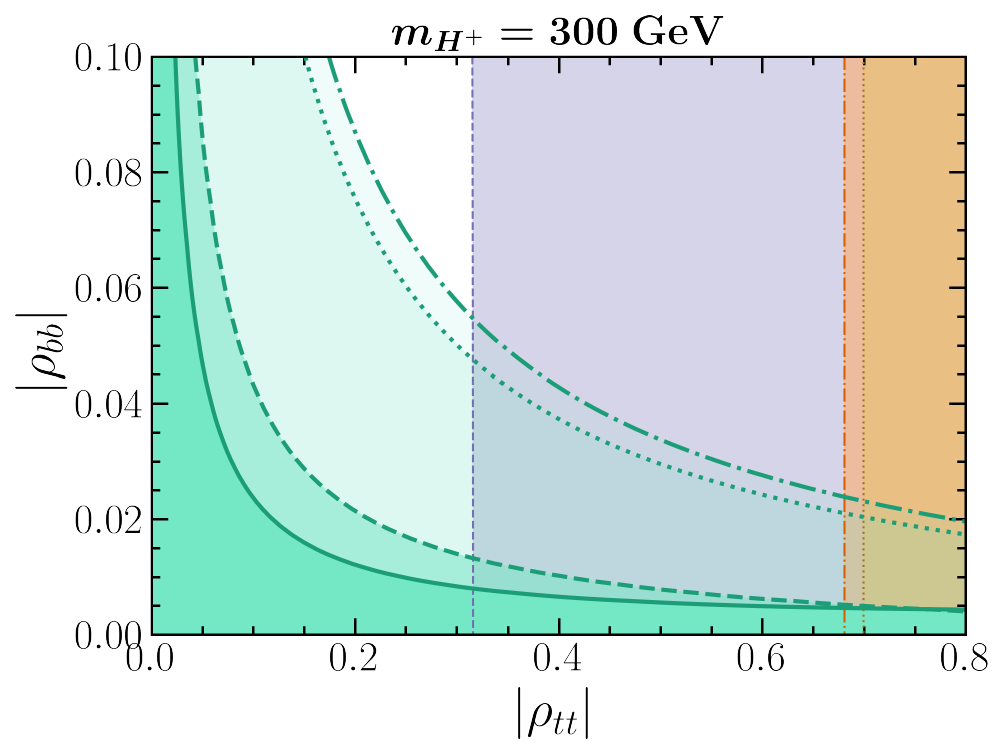}
     \includegraphics[width=0.29\textheight]{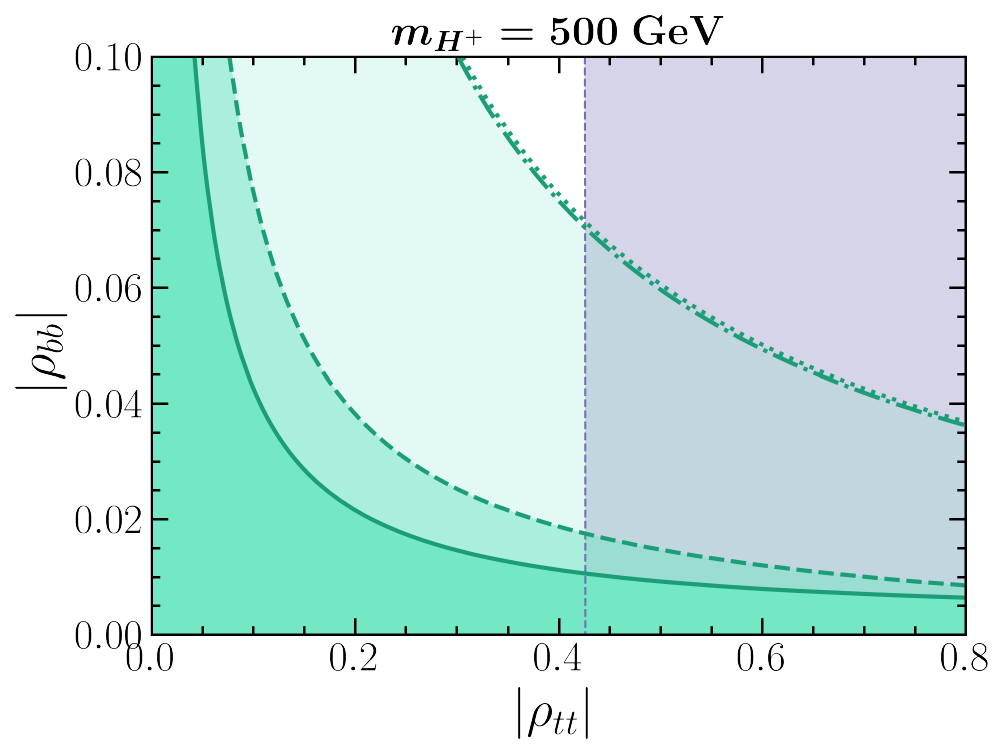}
\vskip-0.1cm
\caption{Results for phase $\phi = 0$ (solid), $\pm \pi/2$ (dot-dash/dots), $\pi$ (dash),
 {\it allowed region} (green) from combined $b\to s\gamma$ observables, and {\it ruled
 out regions} from $B_s$ mixing (purple) and $B_s \to \mu\mu$ (orange).}
\end{figure}
\noindent{\it Combined $b\to s\gamma$ Observables.---}
The impact of current ${\cal B}(B \to X_s\gamma)$ is plotted in Fig.~2 at LO.\footnote{
Things appear quite different at NLO, implying potential future work; difference for 
$\phi = \pm\pi/2$ is also larger at NLO.} 
We see that the curves for $\phi = 0, \pi$ (and also for $\pm\pi/2$) are not so different.
Of some interest is that, if $|\rho_{tt}|$ turns out rather small, 
it could be $\rho_{bb} \sim 0.1$ that drives EWBG, which is {\it also} probed 
by the chiral enhancement effect through Eq.~(6), as one can see from Fig.~2 
for $|\rho_{tt}| < 0.05$.

%
\noindent{\it $\Delta A_{\rm CP}$ Present and Future.---}
More promising may be the CPV observable $\Delta A_{\rm CP}$, Eq.~(7), the 
CPV difference of inclusive $B^+$ vs $B^0$ radiative decays, which we plot in 
Fig.~3. Only the {\it null} sensitivity is shown for maximal phase $\phi = \pm\pi/2$, 
hence we look forward to the actual measurement!

Not knowing what $\Delta A_{\rm CP}$ values might turn up, we explore Eq.~(2), 
the allowed $|\rho_{tt}|$ values for {\it e}EDM cancellation, but set maximal phase 
$\phi = \pm\pi/2$ and show in Table~1. Comparing with the Table above Fig.~2, 
we see that for $|\rho_{tt}| \simeq 0.42$ with maximal phase $\phi = \pm\pi/2$, there 
could be an almost 3$\sigma$ effect! Especially for the case where {\it e}EDM emerges 
real soon, whether one has an echo from {\it n}EDM or not, if a 3$\sigma$ effect 
emerges with full Belle II data, one may face the question of ``To B-III or Not to B-III?'', 
i.e. whether to probe CPV further in $b \to s\gamma$. 

\vskip0.15cm
\begin{table}[h]
\centering
  \begin{tabular}{lcccc}
  \hline\hline
  ~~~$m_{H^+}$ & & {$|\rho_{tt}|$=x, $|\rho_{bb}|$=0.02} &
  & $|\rho_{tt}|$=0.05, $|\rho_{bb}|$=0.1 \\
\hline
    & $x = 0.1 \sqrt{2}$ & $x = 0.2 \sqrt{2}$ & $x = 0.3 \sqrt{2}$ &   \\
    \hline
    $300$ GeV & $\mp(3.041\pm0.046) $ & $\mp(6.026\pm0.091)$ & $\mp(8.902\pm0.134$)
    &  $\mp(5.352\pm0.080$)\\
     $500$ GeV & $\mp(2.055\pm0.031)$ & $\mp(4.097\pm0.063)$ & $\mp(6.111\pm0.093)$
    &  $\mp(3.628\pm0.055)$\\
     \hline\hline
  \end{tabular}
  \caption{In units of $10^{-3}$, for maximal phse $\phi = \phi_{tt} + \phi_{bb} = \pm\pi/2$.}
\end{table}

On the other hand, if $\Delta A_{\rm CP}$ stays consistent with zero with full Belle~II 
data,  it could be pointing at $\Delta A_{\rm CP} \sim 0$, which may support the GUT 
scenario, that the phase of $\rho_{bb}$ cancels against $\rho_{tt}$, as suggested by 
{\it e}EDM cancellation mechanism, and the usual view that $\ell$ and $d$-type quarks 
are grouped together under GUTs.

\begin{figure}[t]
     \includegraphics[width=0.29\textheight]{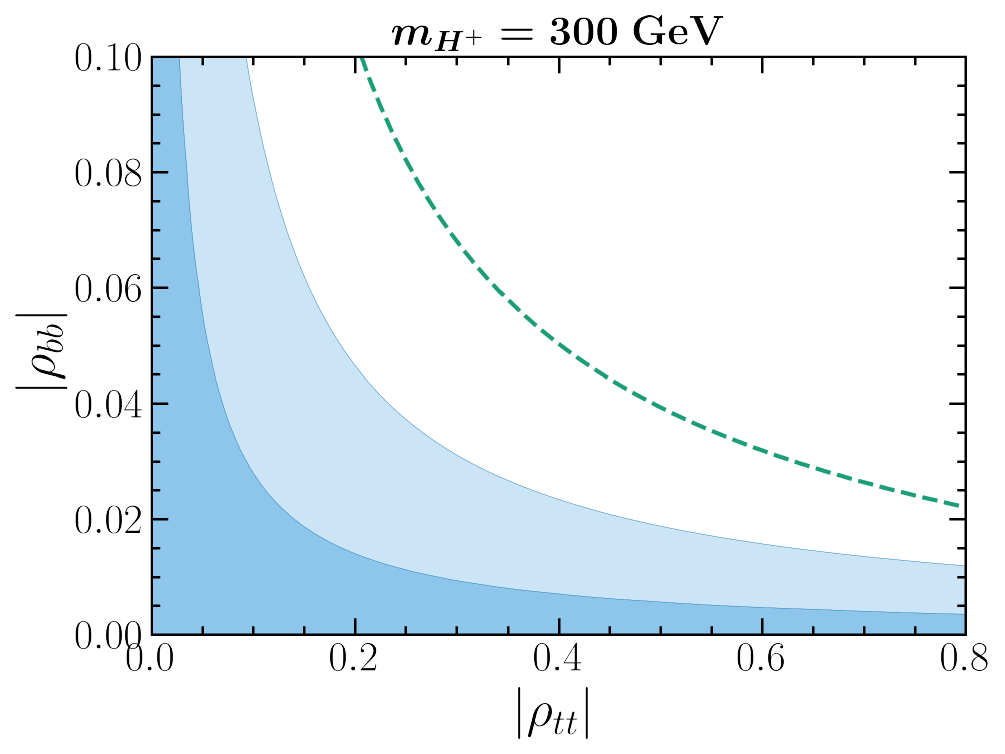}
     \includegraphics[width=0.29\textheight]{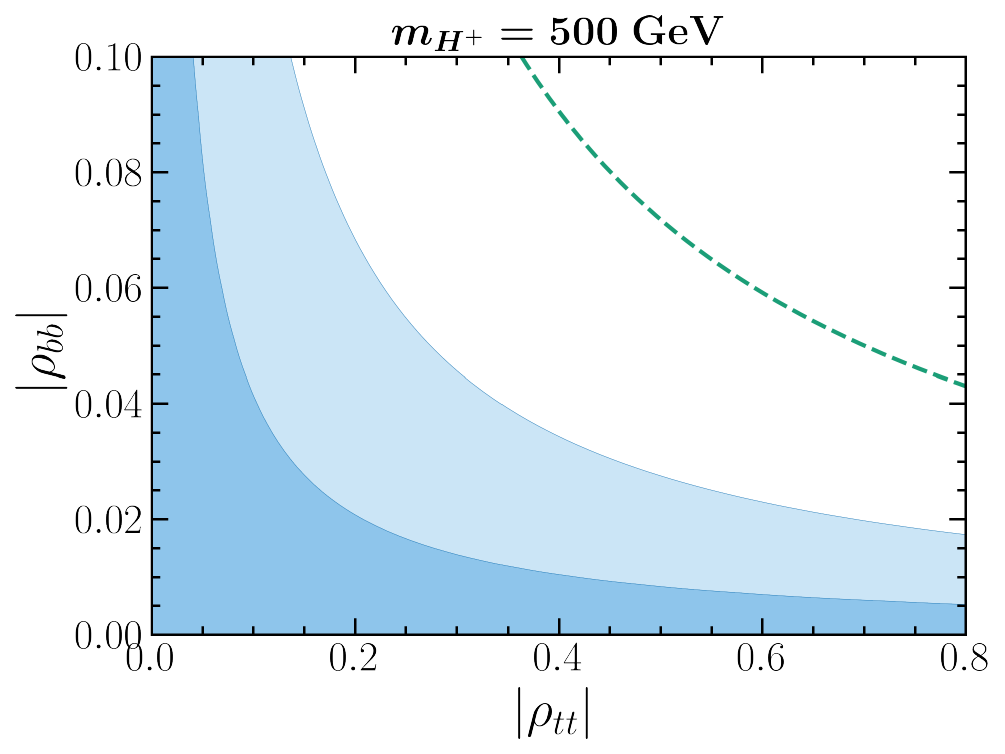}
\vskip-0.1cm
\caption{For $\phi=\pm \pi/2$, (light) blue shaded region {\it allowed} by null $\Delta
 A_{\rm CP}$ with (5) $50\ {\rm ab}^{-1}$ Belle II data, with green dash-dot line the
 individual ${\cal B}(B \to X_s \gamma)$ lower bound. All are $1\sigma$ constraints.
}
\end{figure}

\section{Summary}

Rather than accepting the ``${\cal NNP}$'' fate, we advocate {\it g}2HDM where 
the exotic $H$, $A$ and $H^+$ scalars could well be ${\cal O}(500)$~GeV in 
mass,  responsible for EWBG while accommodating $e$EDM --- and could be 
verified at the LHC, which would be {\it fantastic}! 
Thrown in as bonus would be a bunch of FPCP processes~\cite{Hou:2020itz} .

In this talk we covered only the $b \to s\gamma$ process, the well known probe 
of $H^+$ in SUSY-type 2HDM-II. We illustrate the power of $b \to s\gamma$ 
to probe both $\rho_{tt}$- and $\rho_{bb}$-EWBG through {\it chiral 
enhancement}. Currently, the inclusive rate is the best measured, but we advocate 
that Belle~II can measure $\Delta A_{\rm CP}$ in the future to possibly provide 
a crosscheck on $e$EDM and EWBG!

\vskip0.2cm
\noindent {\bf Acknowledgements.} We thank Marzia Bordone, Fionn Bishop,
 Chen Chen and Kam-Biu Luk for comments and fruitful discussions. We also
 thank Girish Kumar for carefully reading the manuscript.

\end{document}